# Watt-class CMOS-compatible power amplifier

Neetesh Singh[1*], Jan Lorenzen[1], Kai Wang[2], Mahmoud A. Gaafar[1], Milan Sinobad[1], Henry Francis[3], Marvin Edelmann[1], Michael Geiselmann[3], Tobias Herr[1], Sonia M Garcia-Blanco[2], and Franz X. Kärtner[1]

[1]*Center for Free-Electron Laser Science CFEL, Deutsches Elektronen-Synchrotron DESY, Germany*
[2]*Integrated Optical Systems, MESA+ Institute for Nanotechnology, University of Twente, 7500AE, Enschede, The Netherlands*
[3]*LIGENTEC SA, EPFL Innovation Par L, Chemin de la Dent-d'Oche 1B, Switzerland CH-1024 Ecublens, Switzerland*
*neetesh.singh@desy.de

**Abstract:** Power amplifier is becoming a critical component for integrated photonics as the integrated devices try to carve out a niche in the world of real-world applications of photonics. That is because the signal generated from an integrated device severely lacks in power which is due mainly to the small size which, although gives size and weight advantage, limits the energy storage capacity of an integrated device due to the small volume, causing it to rely on its bench-top counterpart for signal amplification downstream. Therefore, an integrated high-power signal booster can play a major role by replacing these large solid-state and fiber-based benchtop systems. For decades, large mode area (LMA) technology has played a disruptive role by increasing the signal power and energy by orders of magnitude in the fiber-based lasers and amplifiers. Thanks to the capability of LMA fiber to support significantly larger optical modes the energy storage and handling capability has significantly increased. Such an LMA device on an integrated platform can play an important role for high power applications. In this work, we demonstrate LMA waveguide based CMOS compatible watt-class power amplifier with an on-chip output power reaching ~ 1W within a footprint of ~4mm$^2$. The power achieved is comparable and even surpasses many fiber-based amplifiers. We believe this work opens up opportunities for integrated photonics to find real world application on-par with its benchtop counterpart.

**Introduction:** Power amplifiers are usually associated with the solid-state and fiber-based bench top systems. That is due to the large energy storage capacity of such systems, owing to the large optical mode and gain area and the long cavity length. Power amplifiers find variety of applications such as in amplifying low noise mode-locked lasers, CW lasers, high power optical frequency comb generation, spectroscopy, laser detection and ranging, material processing and medical applications to name a few [1-14]. Benchtop high power amplifiers have been quite successful for several years, but as we move towards system level miniaturization [9-12], and applications in hostile environments – such as deep space, their size and weight are a major roadblock as they are hard to scale down and mass produce [9-18].

Together with the fiber amplifiers, another technology i.e. the semiconductor amplifiers have been quite successful in the telecommunication industry. Semiconductor power amplifiers, and in particular slab coupled optical waveguide amplifier have shown watt-level amplification in the telecom window [19]; however, so far their photonic integration has met with challenges. There are semiconductor amplifiers that can either be directly mounted or heterogeneously integrated. However, not only the amplified power is merely in the 10s of mW range due to high thermal instability and nonlinear absorption loss [20-23], yield and integration cost is still a concern, especially with silicon photonics as the lattice mismatch between silicon and III-V hinders epitaxial growth, requiring expensive techniques such as flip-chip and wafer-level bonding [21, 23].

Rare-earth gain ions, thanks to the shielding of the 4f shell, are relatively unaffected by their host environment and exhibits rich optical spectra while being barely affected from thermal instabilities and nonlinear losses. Hence, it's no wonder most of the high power solid-state and fiber amplifiers and lasers are based on rare-earth gain medium. In the last two decades, rare-earth gain medium has been satisfactorily demonstrated at chip scale [24-38]. However, the power has remained very low with such devices. Very recently amplification up to 145 mW in the C-band with very long low-loss erbium doped silicon nitride waveguide was demonstrated [38]. However, that was at the expense of a complex fabrication process, and there are still concerns about instabilities from the nonlinear effects (just like in the fibers due to the long lengths and the tight mode confinement), and multi-modedness which can lead to loss of energy to unwanted modes [5].

For high power application it is desirable to have a shorter device that supports large optical modes and active gain region [39-42], which increase the gain saturation power and helps to reduce the nonlinearity while at the same time increase the energy storage capacity. In integrated photonics, however, the very property of tight mode confinement, that enables a small form factor, becomes an impediment to high energy applications, limiting the power to a few 10s of mW [21, 22], and only a few mWs in the mid-infrared window [43].

We have recently explored CMOS compatible large-mode-area (LMA) waveguide for integrated photonics [44, 45]. Such an LMA waveguide helps to increase the energy storage capacity and gain saturation power while incurring negligible nonlinear instabilities, and all that within a compact footprint. In this work, we leverage on such an LMA technology and demonstrate for the first time a CMOS compatible *watt*-class power amplifier with signal amplification reaching up to ~1 W in a compact footprint of 4 mm$^2$ (0.2x21 mm). The signal power reaches and even surpasses the level enjoyed by many commercial bench top fiber amplifiers [46, 47]. The LMA device supports mode area in the range of 10s of µm² increasing the gain saturation power comparable to fiber amplifiers. Moreover, unlike an LMA fiber amplifier, our LMA device supports only fundamental mode propagation, can have tight bends, can be easily interfaced with different photonic components downlink, and allows high pump and signal mode overlap even when they are spectrally far apart.

**Design and experiments:** The gain waveguide cross-section is shown in Fig.1a. The gain waveguide consists mainly of a bottom silicon nitride (SiN) layer buried in silica and a top gain layer. The SiN thickness can vary depending on the wavelength of interest and the foundry of fabrication, allowing flexible fabrication at different CMOS foundries. For the proof of principle demonstration, we chose thulium doped aluminum oxide gain medium ($Tm^{3+}:Al_2O_3$) due to ease in availability and various mid-infrared medical and defense applications [48-53]. The thickness of the gain layer is 1.35 µm on top of a silica cladding layer within which a 800 nm thick SiN layer is buried (such a thick SiN allows seamless integration to conventional nonlinear photonics components). The SiN layer was designed to have a width (w) and the interlayer oxide thickness (g) of around 280 nm and 310 nm, respectively. The simulated TM mode profile of the signal is shown in Fig. 1b, having a pump and signal mode overlap of >98%. The mode area is ~ 30 µm² around the signal (1.85 µm) and the pump (1.61 µm) wavelength. The LMA amplifier design is shown in Fig. 1c. The LMA gain sections are the straight sections within the gain deposited region (enclosed with the green box). To obtain a compact footprint these gain sections are interconnected with each other with the help of tight bends in which the large modes from the LMA sections are transitioned into the small modes (~1.5 µm² mode area) through adiabatic tapers, shown as brown lines. The length of the gain section is ~ 6.2 cm. For pumping the gain medium, we chose in-band pumping scheme. Such a scheme helps to reduce quantum defect (the difference in the pump and the signal photon energy), thus helps to improve the conversion efficiency [54-56] and allows for a relaxed choice of pump wavelength which can be within the range between 1.55 µm to 1.7 µm for thulium doped aluminum oxide. A simplified energy diagram is shown in Fig.1 d showing the upper and lower energy level manifolds. The signal generation can span from 1750 nm to beyond 2 µm [54], thanks to the broadband gain bandwidth of thulium doped glass medium.

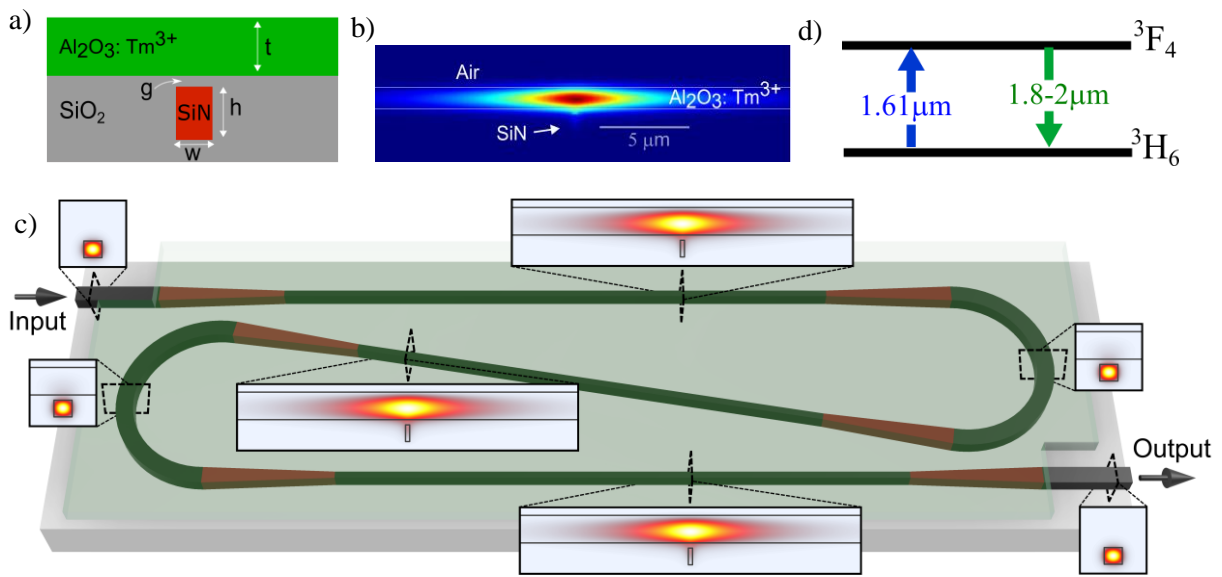

Fig.1 a) The LMA waveguide cross-section, where *t* is the thickness of the gain film (1.35 µm), *g* is the thickness of the interlayer oxide (310 nm), *h* and *w* are the height (800 nm) and width (280 nm) of the SiN layer. b) The signal mode profile at 1.85 µm, with an $A_{eff}$ ~ 30 µm². c) The amplifier schematic, in which the pump and the signal are launched from the input side and the amplified signal is collected at the output. The images of small modes around the bends indicate the tight confinement region where the mode is well confined within the SiN layer, images of large mode indicate the region of LMA where the pump excites the gain ions to upper state to amplify the co-propagating signal. Brown sections indicate the adiabatic tapers to transition the large modes back to the tight modes to allow for the tighter bends. d) A simplified energy diagram of thulium doped alumina with the pump at 1.61 µm and the signal ranging from 1.8 to 2 µm.

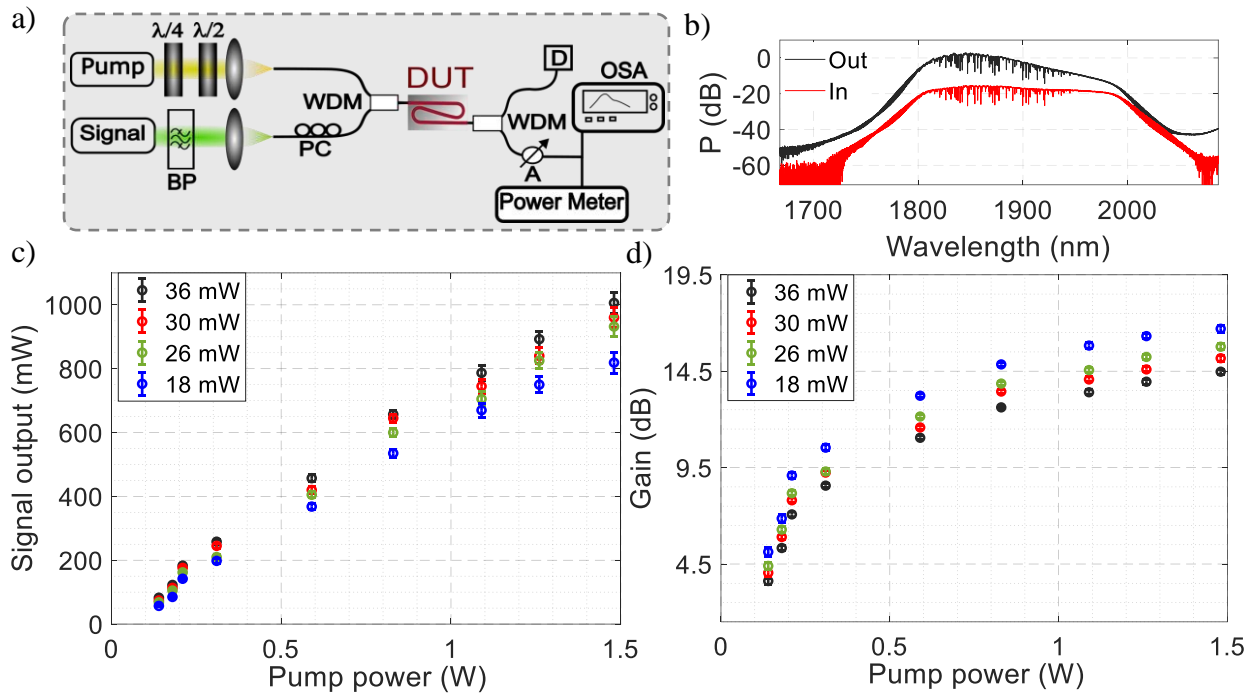

Fig.2. a) The experimental setup. BP is a bandpass filter, PC is a polarization controller, WDM is a wavelength division multiplexer, D is a pump dump, A is a variable attenuator and DUT is the device under test. b) The spectra of the input signal and the amplified output signal. c) On-chip output signal power as a function of pump power for on-chip signal power ranging from 18 mW to 36 mW. d) The net on-chip gain as a function of pump power.

The device was fabricated in a silicon photonics foundry (*Ligentec*) on silicon-nitride-on-silicon platform. The thickness of the SiN layer is 800 nm (thinner layers that are commonly available in CMOS facility can equally be applied with minor modification of the waveguide cross-section). The photonic stack consists of a layer of silicon, bottom silicon dioxide, silicon nitride (800 nm thick), top silicon dioxide and the gain aluminum oxide layer. A gain layer ($Tm^{3+}$:$Al_2O_3$) of 1.35 µm thickness is deposited with radio-frequency (RF) sputtering tool (AJA ATC 15000) at a rate of 5 nm/min at 400°C (method) [57]. The estimated concentration was between 5.5 to 6.5 x$10^{20}$/$cm^3$ and the passive film loss was ≤ 0.1dB/cm at 1.61 µm. Subsequently, the device was tested with the pump at 1.61 µm (a CW laser around 1.61 µm, Alnair labs, TLG 220, amplified with an IPG amplifier, EAR-10-1610-LP-SF). The experimental setup is shown in Fig.2a. The laser was coupled through a lens into the pump channel of the fiber WDM (Thorlab WD1520) which was fusion spliced to a lensed fiber having 3 µm spot size. The signal was around 1.85 µm filtered from a supercontinuum source (NKT Fiannium) with a bandpass filter (Thorlab FB 1900-200) which was launched into the signal channel of the WDM. The coupling loss for the signal and the pump was between 2.5-2.8 dB/facet and 3.5 dB/facet, respectively. We note that better coupling for both signal and pump can be achieved with counter-propagation setup allowing separate alignment for pump and signal, without compromising on signal amplification, as in power amplification the pumping scheme has little affect since the input signal is always strong enough to dominate the amplified spontaneous emission (ASE) [58]. The chip was mounted on a thermally conductive tape for stable operation at high power. The signal was collected at the output with a lensed fibre connected to a WDM which was, down the line, connected through an attenuator to a calibrated optical spectrum analyzer (Yokogawa AQ6376) and a power meter (Thorlab S148C). The on-chip amplified signal power at the output and the net gain as a function of pump power are shown in Fig. 2c, and d, respectively. In this report, the stated optical power is always the on-chip power, unless otherwise stated. Output signal power reached close to 1 W amounting to ~14.5 dB net gain for an input signal power of ~36 mW (maxium signal limited by the source). Upto 16.5 dB net gain was seen for signal around 18 mW before parasitic lasing occurred as shown later. The conversion efficiency at the maximum output ranges between 63-66% (taking into account uncertainty in the pump coupling).

The amplifier was also tested at lower signal power as shown in Fig. 3. The signal ranged from <0.1 mW to ~ 36 mW and the pump was varied from 140 mW to 1.48 W. The green shaded region depicts the parasitic lasing region from the facet reflection which caused gain clamping. This is mainly because the signal amplification overcomes the roundtrip loss leading to lasing, such phenomenon is a well-known nuisance in amplifiers [38, 40, 59].

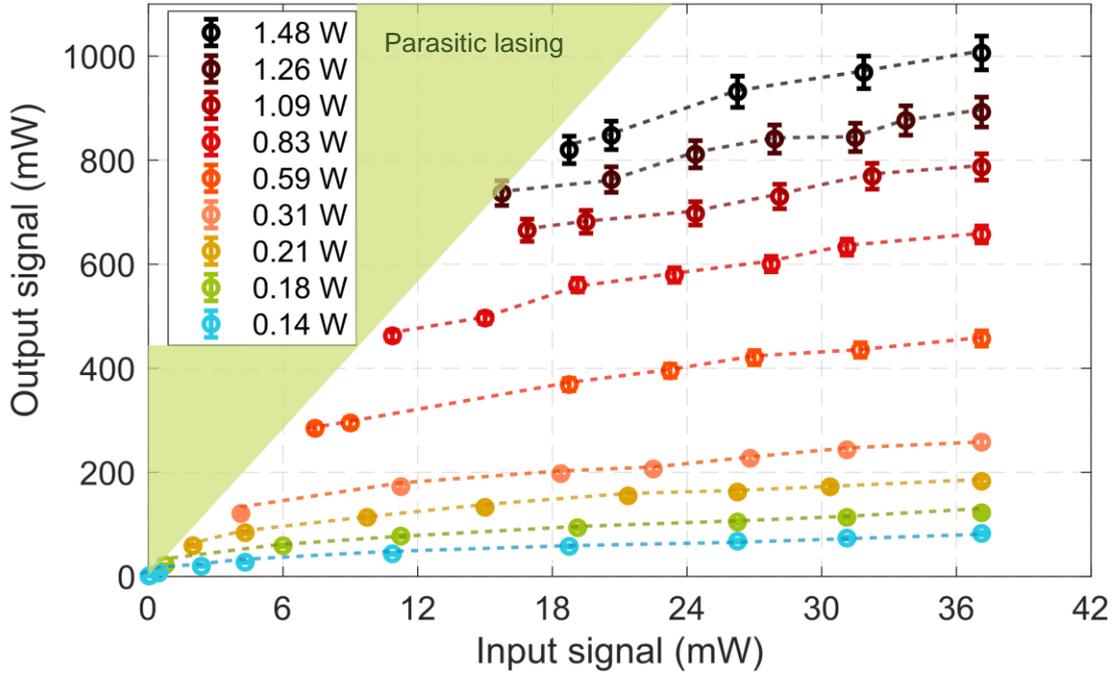

Fig.3. On-chip output signal power as a function of on-chip input signal power at different pump powers. The green shadow section depicts the region of parasitic lasing. The dashed lines that connect the data points are guide for the eye.

The maximum net gain achieved was around 16.5 dB before the parasitic lasing set in, which can be avoided with angled inverse tapers (in a standalone device) or can be avoided altogether in a photonic circuitry where the amplifier will be seamlessly connected to other photonic components without strong Fresnel reflection. By plotting the net gain with respect to the signal power (not shown) the gain saturation power can be estimated to be between 50-80 mW (signal power for which the gain drops to half of its small-signal value), which corresponds to an approximate saturation energy of 35-60 µJ for an upperstate lifetime of 700 µs.

To study the luminescence properties of the active ions in the gain film we perform photoluminescence (PL) study. In order to avoid effects such as, reabsorption, amplified spontaneous emission (ASE) and wavelength dependent loss of the waveguide, we coupled the pump into the waveguide with end-fire coupling and collected the PL with an out-of-plane collection setup (method). The measured upperstate lifetime for the TM mode is shown in Fig. 4a, which exhibits a double exponential decay curve with the $1/e$ point for the lowest pump power being around 1 ms (the time taken for the PL signal to drop by 63.2%). This is much smaller than the expected radiative lifetime of several milliseconds of thulium ions, and can be attributed to the non-radiative phonon based fast decay process [54]. The lifetime drops even further with higher pump power which can be attributed to the increase in decay rate associated with energy transfer up-conversion [54,60-62], which saturates around 720µs. Similar measurements were performed for the signal and pump in TE mode having a larger mode area (<60 µm$^2$), as shown in Fig. 4b. Slightly shorter lifetime was measured for the larger TE mode, ~ 660µs. We also compared the PL strength of the two modes (Fig. 4c), and we observe the larger mode gave stronger PL, which is mainly due to exciting larger number of ions, suggesting a higher gain for the larger mode (we note that the peak of the PL is around 1830 nm unlike the gain (at 1850 nm), that is mainly due to lower absorption loss at 1850 nm). The shape of the PL spectra of the TE and TM mode remained the same except varying in strength for the measured pump power. The mode profiles for the two modes, TE and TM, are shown in Fig.4d, and the pump and the signal mode overlap is over 98% for both of the polarizations. The gain was measured for the TE mode and the difference in gain between the TE and TM mode at different pump powers is shown in Fig. 4d (for the on-chip signal power of ~ 18 mW). The TE mode gain gradually increases with the pump power and surpasses the TM gain around 0.4 W of pump power. At higher pump power TE mode experienced parasitic lasing, which is mainly due to the high gain for the TE mode, which is related to strong photoluminescence for the TE and the fact that a larger optical mode leads to higher gain saturation power; but the effect of stronger interaction with the inhomogeneities in the film cannot be ruled out (causing localized index variation and lifetime variation between TE and TM modes) [63, 64]), which can enhance backscattering and induce early onset of parasitic lasing [65], which can be avoided with the improvement in the film quality which is sensitive to initial deposition conditions [63]. Nevertheless, this demonstrates that even larger mode area reaching the level of a single mode fiber (*SMF-28*) is possible without incurring high losses, thus a smaller amplifier with high energy storage is foreseeable and will be the subject of device optimization in the future studies.

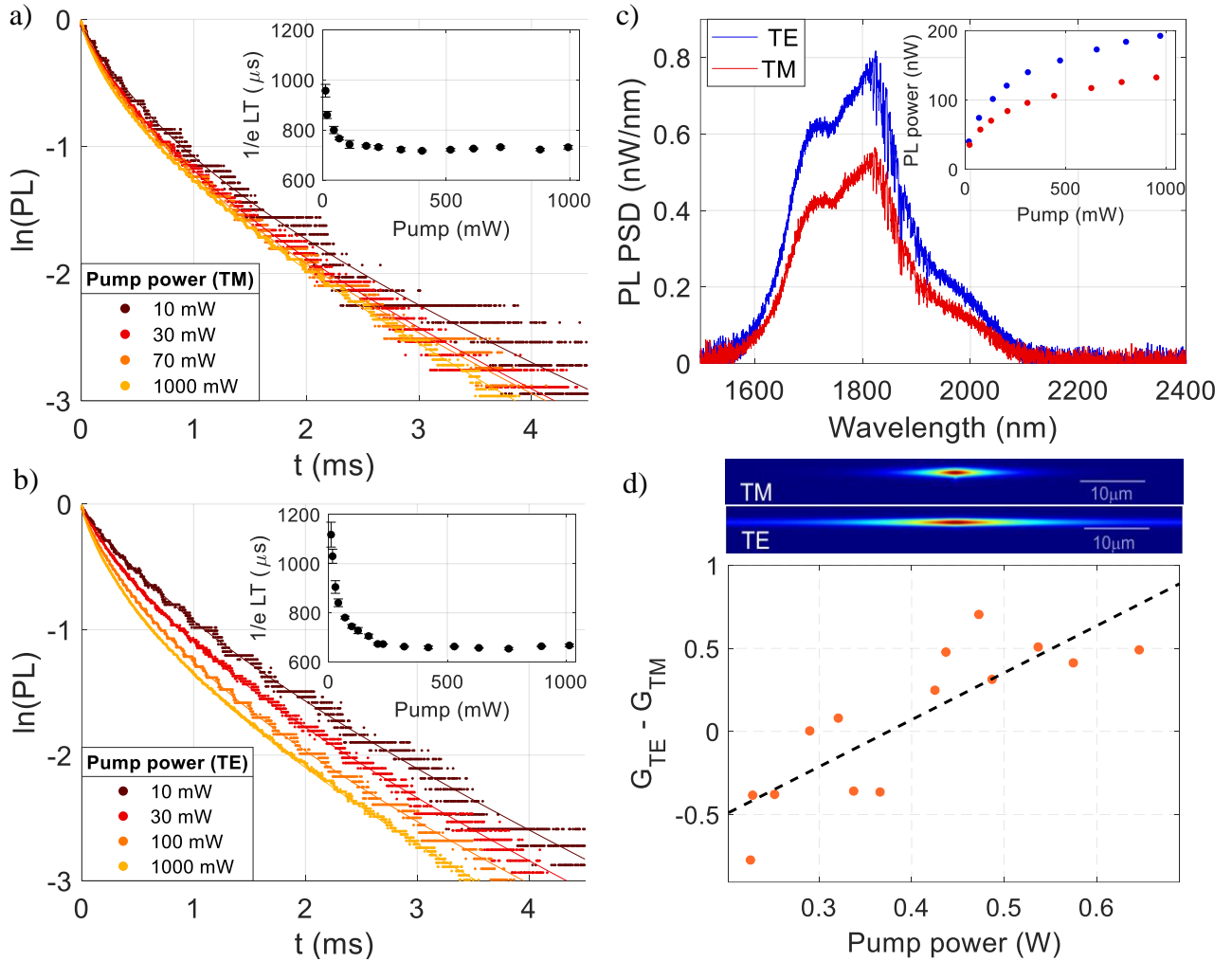

Fig.4. a), and b) Upperstate lifetime measurement of the gain waveguide for the TM mode (30 $\mu m^2$) and the larger TE mode (60 $\mu m^2$) for different pump power. The vertical axis is the natural log of the PL intensity. In the inset, the *1/e* lifetime (LT) with respect to the pump power is shown. c) The PL power spectral density (PSD) curves of the TE and TM modes taken at 950 mW pump power, along with the integrated PL power with respect to the pump power (inset). d) The mode profiles of the TM mode and the TE mode, and the difference between the measured net gain between the two (the straight dashed-line is the linear fit to the data).

Next, we measured the relative intensity noise (RIN) introduced by the amplifier with a signal-source-analyzers (which was locked at the repetition rate of the signal, 78 MHz), see method for details. The noise spectrum at the offset frequency ~ 0.8 kHz and beyond from the carrier is shown in Fig.5. We launched the signal with power >30 mW and varied the pump power from 180 mW to 1.1W. The input signal noise is already high as we use filtered long wavelength edge of the supercontinuum, which is known to have high amplitude noise due to amplified input shot noise through nonlinear processes and spontaneous Raman scattering from a PCF fiber [66]. From a few kHz to several hundred kHz the RIN drops below the input noise, which is due to the high pass filtering effect of the gain with respect to the signal noise [67-69]. This is because the gain is unable to react to the amplitude modulation in the signal (due to noise) below the cut-off frequency (which increases linearly with the output signal power up to the point where the residual pump power starts to increase). In Fig. 5a we see that the curve is shifting to higher cut-off frequencies with the input pump power up to the point where residual pump power at the output starts to increase which is around > 0.85 W of pump power. That is because as the residual pump power increases, the low frequency component of noise from the pump starts to dominate because the gain (population inversion) acts as a low pass filter for the pump fluctuation. Due the same reason in the sub-kHz range (not shown) the noise is dominated by the pump amplitude noise. We have also measured the RIN at different signal power (above parasitic lasing threshold) at a fixed pump power, as shown in Fig.5b. The noise increased from high power to low power input signal by <0.4 dB, which is due to the gradual increase in ASE power as the signal power drops. The noise performance is better than a clad pumped LMA fiber power amplifier for a similar gain system, most likely because our device operates with single fundamental mode, avoiding excitation of ASE in the higher order modes as is the case with LMA fiber amplifiers [40, 70].

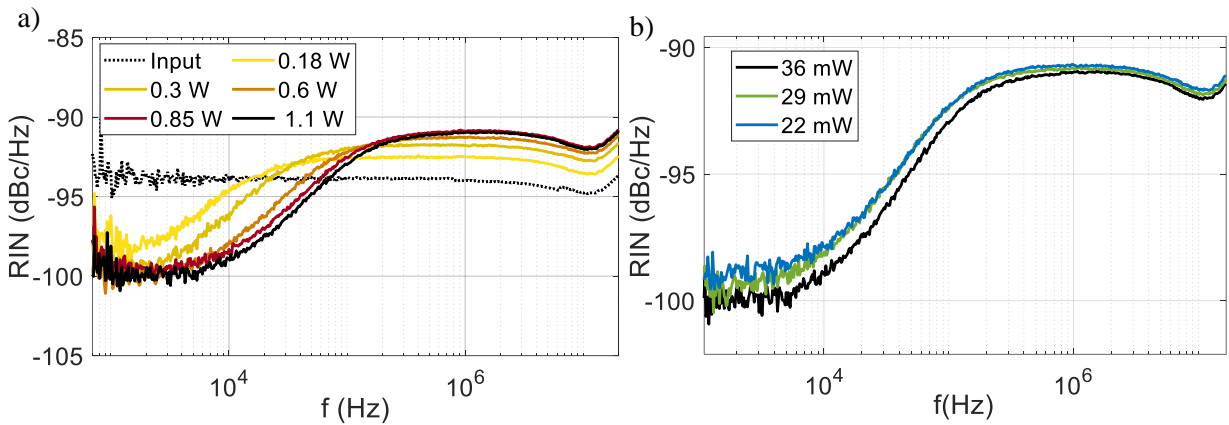

Fig.5 a) The RIN of the signal at different pump powers. The dashed curve is of the reference signal. b) The RIN of the signal for different input signal power at a fixed pump power of 1.1 W.

In conclusion, we have demonstrated the first power amplifier on a CMOS compatible integrated photonics platform with watt-level output power, all that within a footprint of ~4mm$^2$. The LMA waveguide amplifiers demonstrated here is a broadband device; as such, it can be adapted for different spectral windows than can span over an octave without taking a penalty on pump and signal mode overlap. The mode area exploited here for the power amplification was around 30 µm$^2$ and even larger modes (60 µm$^2$) were demonstrated with low loss, suggesting that even more compact amplifiers can be made without comprising on energy storage capacity. This work paves the way forward for signal amplification up to several watts level, in average power, and in energy over several microjoules, which will help mitigate the reliance of integrated photonics on the benchtop systems.

**Acknowledgements:**
This work is supported by EU Horizon 2020 Framework Programme - Grant Agreement No.: 965124 (FEMTOCHIP), and Deutsche Forschungsgemeinschaft (SP2111) contract number PACE:Ka908/10-1.

**Competing interest:**
The authors declare no competing interests.

**Methods:**
**Fabrication:** The photonic stack as mentioned above consists of a layer of silicon, bottom silicon dioxide, silicon nitride, top silicon dioxide, aluminum oxide. The sidewall angle of the etched SiN waveguide is measured by Ligentec to be 89° and the standard variation in SiN thickness and refractive index was ±5%, and ±0.25%, respectively. Silicon substrate thickness = 230 µm, bottom oxide thickness = 4 µm and the top oxide thickness = 3.3 µm. Following the patterning of the SiN layer a silica layer was deposited which was subsequently etched away in the gain region (Fig.1c) upto the point where the silica was 310nm thick on top of the SiN layer - this is the interlayer oxide. SiN fill patterns were fabricated to maintain a high enough density of SiN (>20%) across the reticle to avoid fabrication complications. The gain layer was deposited at the University of Twente with an RF sputtering tool. The chip was mounted in a holder and is loaded into an AJA ATC 15000 RF reactive co-sputtering system through a load-lock and is placed on a rotating holder in the main reaction chamber. A two-inch aluminum target (99.9995% purity) and thulium target are powered through their own RF sources. A power of 200 W is used on the aluminum target and 21 W is used on the thulium target which determines the ion concentration in the film. The deposition temperature was around ~400C and the rate of deposition was around 4 to 5 nm/min which varies by ± 1nm from run to run. After the deposition the samples film quality was characterized with a prism coupling tool (Metricon 2010/M).

**Spectroscopic measurements:** For the upperstate lifetime measurement, the device was optically pumped by an amplified low noise CW laser (Alnair labs, TLG 220). We used a high-power polarization maintaining L-band amplifier (IPG EAR-10-1610-LP-SF). The pump laser goes through a free space setup (consisting of two collimating lenses with a 20 cm free space in between and a chopper was placed in between). The pump was edge coupled to the waveguide and a multimode fiber (Thorlab - M43LO2) coupled light out with an out-of-plane setup. Subsequently, the signal goes through a free space setup: with a lens collimator, bandpass filter (Thorlab FB 1900-200) to remove the residual pump, focusing lens on to a fixed gain amplified InGaAs detector ( THorlab PDA-10D2), which went to an oscilloscope (RS pro RSDS 1304 CFL) through a 20 KHz low noise low pass filter (Thorlab EF-120). For measuring the PL light the pump was launched into the gain layer and the out-of-plane multimode fibre collected the PL light which went straight into the OSA. We note that the PL spectrum was also measured with the 790 nm pump (not shown), and the spectrum was similar to the one obtained with 1.61 µm pumping.

**RIN measurement**: The signal and the pump were coupled into the chip as shown in Fig.1. At the output the signal port of the WDM goes through a free space setup (two collimating lenses coupled to single mode fibers with a 20 cm long free space in between the lenses) in which a bandpass filter to remove pump light was placed. The signal was further attenuated with a variable attenuator to avoid detector saturation (Thorlab VOA-50FC/APC). The light was detected with an InGaAs 12 GHz detector (EOT ET 5000 F/APC), which goes through a bandpass filter (41 – 120 MHz), to detect the signal around 78 MHz (repetition rate of the NKT signal source), which goes through a low noise amplifier ZX60-33LN-S+ (mini circuit) and then into a signal source analyzer to measure the AM noise (SSA-E5052B).

**Refractive index:** The optical constants of the $Al_2O_3$ film was measured with VASE ellipsometer (J.A. Woollam Co.) using an ellipsometer operating from 240 nm to beyond 11 µm. To isolate the influence due to the silicon and silica substrates, samples with only oxidized silicon and bare silicon were also measured. The refractive index at 1.9 µm of the film was measured to be ~ 1.7.

**References:**
1. A. E Willner et. al. "Optics and Photonics: Key Enabling Technologies," Proc. IEEE, 100 (2012).
2. Wei Shi, et. al. "Fiber lasers and their applications," Appl. Optics, 53 (2014).
3. D. J. Richardson et. al. "High power fiber lasers: current status and future perspectives," *J. Opt. Soc. Am. B* 27(11), B63–B92 (2010).
4. M. N. Zervas "High Power Fiber Lasers: A Review," IEEE J Sel Top Quantum Electron IEEE J SEL TOP QUANT, 20 (2014).
5. C. Jauregui, et. al. "High-power fibre lasers," Nat. Photon. 7 (2013).
6. A. Cingoz et. al., "Direct frequency comb spectroscopy in the extreme ultraviolet," Nature, 482, (2012).
7. T. R. Schibli, et. al." Optical frequency comb with submillihertz linewidth and more than 10 W average power," Nat. Photon. 2, (2008).
8. K. Sugioka and Y. Cheng, "Ultrafast laser – reliable tools for advanced materials processing," Light: Science and Appl. 3, (2014).
9. X. Zhang, et. al. "A large scale microelectromechanical systems bases silicon photonic LiDAR," Nature, 603 (2022).
10. P. D. Haye, et. al."Phase coherent microwave to optical link with a self-referenced microcomb," Nat. Photon. 10, (2016).
11. D. T. Spencer, et. al. "An optical frequency synthesizer using integrated photonics," Nature, 557, (2018).
12. N. Singh et. al. "Silicon photonics optical frequency synthesizer," Laser Photon. Rev. 14, (2020).
13. www.seminex.com "Seminex unveils high gain optical amplifiers for LIDAR" Seminex news (2022). S. Aboujja, et.al. "High performance semiconductor optical amplifier and array for FMCW LiDAR in high speed autonomous vehicles," SPIE LASE, 12403 (2023).
14. www.lumentum.com, E. Canoglu et. al. "Semiconductor lasers and optical amplifiers for LiDAR photonic integrated circuits," International semiconductor laser conference, 21531815 (2021).
15. J. M. Dailey, et. al. "High output power laser transmitter for high efficiency deep space optical communications," SPIE LASE, 109100M (2019).
16. D. M. Cornwell "NASA's optical communications program for 2017 and beyond," IEEE Intern. Conf. Space Optical Syst. Appl., (2017).
17. H. Hemmati et. al.,"Deep-Space Optical Communications: Future Perspectives and Applications," Proc. of IEEE, 99 (2011).
18. H. Kaushal, and G. Kaddoum, "Optical communication in space: challenges and mitigation techniques," IEEE comm. Surver and Tutorials, 19 (2017).
19. P. W. Juodawlkis et. al., "High-Power, Low-Noise 1.5-$\mu$m Slab-Coupled Optical Waveguide (SCOW) Emitters: Physics, Devices, and Applications," IEEE J Sel Top Quantum Electron IEEE J SEL TOP QUANT, 17 (2011).
20. H. Zhao, et. al. "High power indium phosphide photonic integrated circuits," *IEEE J Sel Top Quantum Electron* 25, (2019).
21. M. L. Davenport et. al. Heterogeneous Silicon/III–V Semiconductor Optical Amplifiers," IEEE J Sel Top Quantum Electron IEEE J SEL TOP QUANT, 22 (2016).
22. K. V Gasse, et. al."27 dB gain III-V on silicon semiconductor optical amplifier with >17 dBm output power," *Opt. Express*, 27, 2019.

23. Z. Zhou, B. Yin and J. Michel, "On-chip light sources for silicon photonics", *Light Sci. Appl.*, vol. 5, pp. 1-13, 2015.
24. J. Kenyon, "Erbium in silicon", *Semicond. Sci. Technol.*, vol. 20, pp. R65-R84, 2005.
25. Purnawirman, et. al. "C- and L-band erbium-doped waveguide lasers with wafer-scale silicon nitride cavities," *Opt. Lett*. 38, (2013).
26. L. Agazzi, et. al., "Monolithic integration of erbium-doped amplifiers with silicon-on-insulator waveguides," *Opt. Express* **18**(26), 27703–27711 (2010)
27. M. Belt and D. J. Blumenthal, "High temperature operation of an integrated erbium- doped DBR laser on an ultra-low-loss Si3N4 platform," *Optical Fiber Communication Conference*, OSA Technical Digest (online) (Optical Society of America, 2015), paper Tu2C.7
28. E. S. Magden, N. Li, J. Purnawirman, D. B. Bradley, N. Singh, A. Ruocco, et al., "Monolithically-integrated distributed feedback laser compatible with CMOS processing", *Opt. Express*, vol. 25, no. 15, pp. 18058-18065, 2017.
29. N. Li, D. Vermeulen, Z. Su, E. S. Magden, M. Xin, N. Singh, et al., "Monolithically integrated erbium-doped tunable laser on a CMOS-compatible silicon photonics platform", *Opt. Express*, vol. 26, pp. 16200-11, 2018.
30. J. Rönn, W. Zhang, A. Autere, X. Leroux, L. Pakarinen, C. Alonso-Ramos, et al., "Ultra-high on-chip optical gain in erbium-based hybrid slot waveguides", *Nat. Communications*, vol. 10, pp. 432, 2019.
31. H. Sun, L. Yin, Z. Liu, Y. Zheng, F. Fan, S. Zhao, et al., "Giant optical gain in a single-crystal erbium chloride silicate nanowire", *Nat. Photonics*, vol. 11, pp. 589-593, 2017.
32. A Choudhary, et al., "A diode-pumped 1.5 μm waveguide laser mode-locked at 6.8 GHz by a quantum dot SESAM", *Laser Phys. Lett.*, vol. 10, pp. 1-4, 2013.
33. H. Byun, et al., "Integrated low-jitter 400-MHz femtosecond waveguide laser", *IEEE Photo. Tech. Lett*, vol. 21, pp. 763-765, 2009.
34. J. D. Bradley, et al., "Monolithic erbium- and ytterbium doped microring lasers on silicon chips", *Opt. Express*, vol. 22, pp. 12226-12237, 2014.
35. K. van Dalfsen et. al. "Thulium channel waveguide laser with 1.6 W of output power and ∼80% slope efficiency," *Opt. Lett.* 39, (2014).
36. K. Shtyrkova, et al., "Integrated CMOS-compatible Q-switched mode-locked lasers at 1900 nm with an on-chop artificial saturable absorber", *Optics Express*, vol. 27, no. 3, pp. 3542-3556, 2019.
37. F. X. Kärtner et.al. "Integrated CMOS-Compatible Mode-Locked Lasers and Their Optoelectronic Applications", *Proc. SPIE 10686*, 14 (2018).
38. Y. Liu, et. al. "A photonic integrated circuit based erbium doped amplifier," *Science*, 376, 2022.
39. D. Taverner, et. al. "158-uJ pulses from a single-transverse-mode, large-mode-area erbium-doped fiber amplifier," Opt. Lett., 22, (1997)
40. C. C. Renaud, et. al. " Characteristics of Q-switched cladding-pumped ytterbium-doped fiber lasers with different high-energy fiber designs," IEEE J. Quantum Electronics, 27 (2001)
41. J. Limpert, et. al. "100-W average power, high-energy nanosecond fiber amplifier," Appl. Phys. B, 75 (2002).
42. M. Y. Cheng, et. al." High-energy and high-peak-power nanosecond pulse generation with beam quality control in 200-$\mu$m core highly multimode Yb-doped fiber amplifiers," Opt. Lett, 30 (2005).
43. N. Volet, et. al. "Semiconductor optical amplifier at 2 μm wavelength on silicon," *Laser Photonics Rev.* 11, 2017.
44. N. Singh et.al. "Towards CW modelocked laser on chip – a large mode area and NLI for stretched pulse mode locking", *Opt. Express*, 28, 15 (2020).
45. N. Singh et. al. Chip-scale, CMOS-compatible, high energy passively Q-switched laser, "arXiv preprint arXiv:2303.00849
46. www.advaluephotonics.com. Advalue photonics - 2 Micron Fiber amplifier (Ap-Amp).
47. www.thorlabs.com. PicoLuz – Thulium doped fiber amplifier.
48. K Scholle, et al. "2 μm Laser Sources and Their Possible Applications," *Frontiers in guided wave optics and optoelectronics*, 2010.
49. K. Yang, et. al. "Q-Switched 2 micron solid-state laser and their applications," *Frontiers in guided wave optics and optoelectronics* (2019).
50. C. Boone, "Medical applications are a surgical fit for 2 μm lasers," *Laser Focus World*, 2022
51. X. Xie, et. al. "A brief review of 2 μm laser scalpel," *IEEE 5th Optoelectronics* 2022.
52. U. N. Singh, "Progress on high energy 2 micron solid state laser for NASA space-based wind and carbon dioxide measursments," 011 *IEEE Photonics Society Summer Topical Meeting Series*, 2011.
53. Z. Li. Et. al. "Thulium-doped fiber amplifier for optical communications at 2 μm," *Opt. Express*, 21, (2013).
54. A. Sincore, et. al. "High average power thulium-doped silica fiber lasers: review of systems and concepts,", IEEE J Sel Top Quantum Electron IEEE J SEL TOP QUANT, 24 (2018).
55. M. Lenski et. al. "Inband-pumped, high power thulium doped fiber amplifiers for an ultrafast pulsed operation," Opt. Express, 24 (2022)
56. G. D. Goodno et. al. "Low-phase-noise, single-frequency, single-mode 608 W608 W thulium fiber amplifier," Opt. Lett. 34, 2009
57. C. I. van. Emmerik, et. al. "Relative oxidation state of the target as guideline for depositing optical quality RF reactive magnetron sputtered $Al_2O_3$ layers," Opt. Mat. Express, 10, (2020).
58. P. C. Becker et. al. "Erbium doped fiber amplifiers: fundamentals and technology," Academic press inc. 1999.

59. D. C. Brown et. al. "Parasitic oscillations, absorption, stored energy density and heat density in active-mirror and disk amplifiers," Appl. Optics, 17, (1978).
60. D. A. Simpson, et. al. "Energy transfer up-conversion in Tm3+ - doped silica fiber," J. Non-Crystalline Solids, 352, 2006.
61. G. Nykolak et. al. "Concentration-dependent $^4I_{13/2}$ lifetimes in $Er^{3+}$ doped fibers and $Er^{3+}$ doped planar waveguides," IEEE Photon. Technology Lett. 5 (1993).
62. K. Kuroda et. al. "Pump-probe measurment of metastable state lifetime reduced by cooperative upconversion in a high-concentration erbium doped fiber," Appl. Optics. 57 (2018).
63. R. A. Gardner et. al. "Stability of RF-sputtered alumium oxide", Journal of Vaccum Science and Techn. 14, (1977).
64. F. Hacker, et. al. " RF-sputterd SiO2 films for optical applications," Thin Solid Films, 97 (1982).
65. A.A. Hardy and R. Oron, "Amplified spontaneous emission and Rayleigh backscattering in strongly pumped fiber amplifiers," J. Lightwave. Techn. 16, 1998.
66. K. L. Corwin, et. al. "Fundamental noise limitations to supercontinuum generation in microstructure fiber," Phys. Rev. Lett. 90 (2003).
67. S. Novak and A. Moesle, "Analytic model for gain modulation in EDFAs," J. lightwave Technol. 20, (2002).
68. H. Tunnermann et. al. "Gain dynamics and refractive index changes in fiber amplifiers: a frequency domain approach," Opt. Express. 20 (2012).
69. J. Zhao et. al. "Gain dynamics of clad-pumped Yb-fiber amplifier and intensity noise control," Opt. Express, 25 (2017).
70. G. Guiraud et. al. "High power and low intensity noise laser at 1064nm," Opt. Lett. 41 (2016).